\documentclass[aps,prb,twocolumn,superscriptaddress]{revtex4}

\usepackage{graphicx,color}

\input epsf.sty
\usepackage{graphicx}

\begin{document}

\preprint{\today}


\title{
Resonant inelastic X-ray scattering study of intra-band charge 
excitations in hole-doped high-$T_c$ cuprates
}

\author{Shuichi Wakimoto
}
\affiliation{ Quantum Beam Science Directorate, Japan Atomic Energy Agency,
   Tokai, Ibaraki 319-1195, Japan }

\author{Kenji Ishii}
\affiliation{ Synchrotron Radiation Research Center, Japan Atomic Energy Agency,
   Hyogo 679-5148, Japan }

\author{Hiroyuki Kimura}
\affiliation{ Institute of Multidisciplinary Research for Advanced Materials, 
   Tohoku University, Sendai 980-8577, Japan }

\author{Kazuhiko Ikeuchi}
\affiliation{ Research Center for Neutron Science and Technology, Comprehensive 
   Research Organization for Science and Society,
   Tokai, Ibaraki 319-1195, Japan }

\author{Masahiro Yoshida}
\affiliation{ School of Science and Technology, Kwansei Gakuin Universitry,
   Sanda, Hyogo 669-1337, Japan }

\author{Tadashi Adachi}
\affiliation{ Department of Applied Physics, Tohoku University, 
   Sendai 980-8579, Japan }

\author{Diego Casa}
\affiliation{ CMC-XOR, Advanced Photon Source, Argonne National
   Laboratory, Argonne, Illinois 60439, USA }

\author{Masaki Fujita}
\affiliation{ Institute for Materials Research, Tohoku University, Katahira,
   Sendai 980-8577, Japan }

\author{Yasushi Fukunaga}
\affiliation{ Department of Applied Physics, Tohoku University, 
   Sendai 980-8579, Japan }

\author{Thomas Gog}
\affiliation{ CMC-XOR, Advanced Photon Source, Argonne National
   Laboratory, Argonne, Illinois 60439, USA }

\author{Yoji Koike}
\affiliation{ Department of Applied Physics, Tohoku University, 
   Sendai 980-8579, Japan }

\author{Jun'ichiro Mizuki}
\affiliation{ Synchrotron Radiation Research Center, Japan Atomic Energy Agency,
   Hyogo 679-5148, Japan }
\affiliation{ School of Science and Technology, Kwansei Gakuin Universitry,
   Sanda, Hyogo 669-1337, Japan }

\author{Kazuyoshi Yamada}
\affiliation{ Institute for Materials Research, Tohoku University, Katahira,
   Sendai 980-8577, Japan }
\affiliation{ WPI Research Center, Advanced Institute for Materials Research, 
   Katahira, Sendai 980-8577, Japan}
\affiliation{ Institute of Materials Structure Science, 
   High Energy Accelerator Research Organization, Tsukuba, Ibaraki 305-0801, Japan }

\date{\today}

\begin{abstract}

We have performed resonant inelastic x-ray scattering (RIXS) near the Cu-$K$ 
edge on cuprate superconductors La$_{2-x}$Sr$_{x}$CuO$_{4}$, 
La$_{2-x}$Ba$_{x}$CuO$_{4}$, La$_{2-x}$Sr$_{x}$Cu$_{1-y}$Fe$_{y}$O$_{4}$ 
and Bi$_{1.76}$Pb$_{0.35}$Sr$_{1.89}$CuO$_{6+\delta}$, covering underdoped 
to heavily overdoped regime and focusing on charge excitations inside the 
charge-transfer gap.
RIXS measurements of the 214 systems with $E_i = 8.993$~keV have revealed 
that the RIXS intensity at 1~eV energy transfer has a minimum at $(0,0)$ 
and maxima at $(\pm0.4 \pi, 0)$ and $(0, \pm0.4 \pi)$ for all doping points regardless 
of the stripe ordered state, suggesting that the corresponding structure is 
not directly related to stripe order.
Measurements with $E_i = 9.003$~keV on metallic La$_{1.7}$Sr$_{0.3}$CuO$_{4}$ 
and Bi$_{1.76}$Pb$_{0.35}$Sr$_{1.89}$CuO$_{6+\delta}$ 
exhibit a dispersive intra-band excitation below 4~eV, similar to 
that observed in the electron-doped Nd$_{1.85}$Ce$_{0.15}$CuO$_{4}$.  
This is the first observation of a dispersive intra-band excitation in a 
hole doped system, evidencing that both electron and hole doped systems
have a similar dynamical charge correlation function.

\end{abstract}


\maketitle

\section{Introduction}

Among the many functional physical properties in strongly correlated electron 
systems, high-$T_c$ superconductivity continues to pose long-term challenges and 
attract many researchers.  Elementary excitations in high-$T_c$ superconducting 
cuprates, such as spin and charge excitations, are important subjects 
in condensed matter physics not only for understanding superconductivity 
but also for a more fundamental understanding of strongly correlated electron 
systems.

To date, spin excitations in high-$T_c$ cuprates have been extensively studied by 
neutron scattering~\cite{Birgeneau2006} and more recently by Resonant Inelastic 
X-ray Scattering (RIXS) at Cu-$L_{3}$ edge,~\cite{Braicovich2010,LeTacon2011,Dean2012} 
while charge excitations have been studied by optical 
spectroscopy~\cite{Uchida1991} and RIXS at Cu-$K$ edge.~\cite{Kim2002}
In addition to its ability to determine the energy and momentum of excitations, 
recent RIXS studies at the Cu-$L_3$ edge have demonstrated the ability of RIXS 
to detect very low energy charge fluctuations in the (Y,Nd)Ba$_2$Cu$_3$O$_{6+y}$ 
system.\cite{Ghiringhelli2012}
Moreover, the RIXS technique near the Cu-$K$ edge is sensitive to charge 
excitations triggered by the core hole potential in the intermediate state of the 
RIXS process.  In fact, RIXS near the Cu-$K$ edge
has been successfully utilized to measure charge excitations 
across the charge transfer (CT) gap in various strongly correlated electron 
systems,~\cite{Kim2002,Abbamonte1999,Hasan2000,Lu2005,Ishii2005,Ishii2007}  
with excitation energy typically in 2-4 eV range, depending on the particular system and 
doping levels.
One characteristic feature of the high-$T_c$ cuprates is that the 
continuum-like charge excitation appears inside the CT gap upon hole or 
electron doping.~\cite{Wakimoto2005,Ellis2011}  
Recent technical developments towards finer energy resolution have 
enabled us to study this in-gap charge 
excitations in more detail.

High-$T_c$ cuprates are one of the most suitable systems to study charge 
excitations inside the CT gap since 
it can be approximated by a single-band model, 
and 
therefore making theoretical handling relatively easy. 
Ishii {\it et al}.~\cite{Ishii2005a} have reported a dispersive intra-band excitation 
inside the CT gap in the electron-doped system Nd$_{1.85}$Ce$_{0.15}$CuO$_{4}$.  
Based on comparison with theoretical calculations, they claimed that this 
dispersive excitation is consistent with the dynamical charge-charge correlation 
function $N(q,\omega)$.
On the other hand, such a dispersive feature inside the CT gap has not been 
reported in hole-doped high-$T_c$ cuprates.  Instead, it has been reported that the 
continuum-like excitation of La$_{2-x}$Sr$_{x}$CuO$_{4}$ near 1~eV at 
high-symmetry points increases with doping,~\cite{Wakimoto2005,Ellis2011} 
showing an anomalous increase across the 1/8 doping point.~\cite{Ellis2011}  
Our previous measurements 
showed that the charge excitation near 1~eV is enhanced at the $(\pm\pi/2, 0)$ and 
$(0, \pm\pi/2)$ positions in stripe-ordered La$_{1.875}$Ba$_{0.125}$CuO$_{4}$ 
and La$_{1.88}$Ba$_{0.12}$CuO$_{4}$, while such behavior is absent in 
La$_{1.92}$Ba$_{0.08}$CuO$_{4}$.~\cite{Wakimoto2009}

In order to reconcile these experimental facts, we have carried out RIXS 
measurements near the Cu-$K$ absorption edge using single crystals of 
La$_{2-x}$(Sr,Ba)$_{x}$CuO$_{4}$ (LSCO, LBCO) covering a wide hole concentration 
range from $x=0$ to $x=0.30$, focusing on charge excitations inside the CT gap.  
We used Cu-$K$ edge RIXS since it is more sensitive to charge excitations 
than at other absorption edges.
The measurements were carried out at nearly 90 degrees scattering angle with 
the scattered photon direction parallel to the incident photon polarization.  
This scattering geometry greatly reduces the elastic intensity and enables us 
to study RIXS spectra at low energies ($\sim 1$~eV).  
We have found that the intensity of the charge excitation near 1~eV measured with 
incident photon energy $E_i = 8.993$~keV has maxima around $(\pm0.4 \pi, 0)$ and 
$(0, \pm0.4 \pi)$ for all doped samples, and 
this intensity increases monotonically with doping.  
We have also carried out RIXS measurements with $E_i=9.003$~keV for 
metallic LSCO x=0.30 and Bi$_{1.76}$Pb$_{0.35}$Sr$_{1.89}$CuO$_{6+\delta}$ 
(Bi2201).  By subtracting both the elastic component and high-energy RIXS intensities, 
we have found that the charge excitation below 4~eV has a $q$-dependence 
similar to the intra-band excitation observed in the electron-doped 
Nd$_{1.85}$Ce$_{0.15}$CuO$_{4}$, suggesting that both electron and 
hole doped systems have a similar dynamical charge correlation function.

\section{Experimental procedure}

Compositions used for the present study were La$_{2}$CuO$_{4}$, 
La$_{2-x}$Ba$_{x}$CuO$_{4}$ (LBCO) with $x=0.08$ and 0.125, 
La$_{2-x}$Sr$_{x}$CuO$_{4}$ (LSCO) with $x=0.15$ and 0.30, 
Bi$_{1.76}$Pb$_{0.35}$Sr$_{1.89}$CuO$_{6+\delta}$ (Bi2201), and 
La$_{1.85}$Sr$_{0.16}$Cu$_{0.99}$Fe$_{0.01}$O$_{4}$ (Fe-doped LSCO). 
The last compound shows static incommensurate magnetic and charge 
order induced by the substitution of Cu$^{2+}$ ions with Fe$^{3+}$ 
ions.~\cite{He2011}
All crystals were grown by the travelling-solvent floating-zone method. 
As-grown crystals were cut into disk shapes with the $c$-axis normal to 
the disk.  

RIXS measurements for LSCO, LBCO and Bi2201 were made using the 
spectrometer at BL11XU beamline of SPring-8, and those for Fe-doped 
LSCO were made at the spectrometer at beamline 9-ID-B 
at the Advanced Photon Source, Argonne National Laboratory. 
For the measurements of LSCO and LBCO at BL11XU, the incident photon 
energies were either 8.993 or 9.003~keV.  These  
energies are indicated in Fig.~1~(a) by two dashed vertical lines over 
the fluorescence spectra of LSCO $x=0$ and $0.30$.  Since the fluorescence 
spectrum is not very sensitive to hole-concentration, we have used 
the same incident energies across samples. 
For the measurements of Bi2201 at BL11XU, an incident photon energy of 
9.003~keV was used.  Again, this energy is indicated in Fig.~1~(b) over 
the fluorescence spectrum for Bi2201.
For the measurements of Fe-doped LSCO at 9-ID B, we have used an incident phonon 
energy of 8.994~keV.

\begin{figure}
\includegraphics[width=6cm]{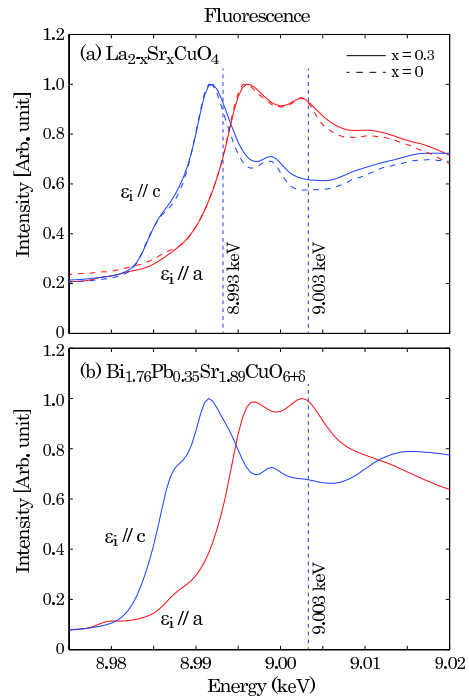}
\caption{(Color online) Fluorescence spectra for (a) La$_{2}$CuO$_{4}$ 
(dashed line) and La$_{1.7}$Sr$_{0.3}$CuO$_{4}$ (solid line) as a 
representative spectra of the LSCO system and 
(b) Bi$_{1.76}$Pb$_{0.35}$Sr$_{1.89}$CuO$_{6+\delta}$.  For 
each material, fluorescence was measured with two conditions: $\epsilon_i 
\parallel c$ and $\epsilon_i \parallel a$.  For the LSCO system, RIXS 
measurements were done at $E_i = 8.993$ and $9.003$~keV, while those for 
Bi$_{1.76}$Pb$_{0.35}$Sr$_{1.89}$CuO$_{6+\delta}$ at $E_i = 9.003$~keV.}
\end{figure}

Scattering geometry used for all RIXS measurements is schematically 
drawn in Fig.~2~(d).  Here, the $(a, c)$ plane in tetragonal notation 
is in the scattering plane.  
Since the compounds we studied are two dimensional systems, we assign
the $(0,0,L)$ position as the $\Gamma$-point $(0, 0)$, and the $(0.5,0,L)$
and $(0.5,0.5,L)$ positions as zone-boundaries $(\pi, 0)$ and $(\pi, \pi)$ 
of the basal CuO$_{2}$ plane, respectively.  
The scattering angle $2\theta$ was kept at 
$\sim 90$~degrees by adjusting the $L$-value so that the scattered photon 
direction is parallel to the incident photon polarization $\epsilon_i$.  
This geometry reduces the elastic component, resulting in better access 
to low energy RIXS signals.  
The above geometry was be naturally achieved with the BL11XU spectrometer since it has a horizontal
scattering arm, parallel to the horizontal $\epsilon_i$.  
The 9-ID B spectrometer has a vertical 
scattering arm instead, so for these measurements we utilized a diamond 
phase plate 
to set $\epsilon_i$ vertical.  
For the experiments at BL11XU (9-ID B), a Si(111) monochromator, 
a secondary Si(400) (Si(444)) channel-cut monochromator, a bent Ge(337) 
analyzer, and a point silicon detector (a MYTHEN strip detector) were used.
For both instruments, the energy resolution was $\sim 350$~meV.

We measured RIXS spectra by scanning the scattered photon energy while the 
incident photon energy was fixed near the Cu-$K$ edge.  
In addition, we measured the momentum transfer ${\rm\bf q}$ dependence of the RIXS 
intensity at fixed energy transfer.  
In this case, the incident beam angle $\theta$ with respect to the sample 
surface varies with momentum transfer.  Following the argument
by Pfalzer {\it et al.},~\cite{Pfalzer1999} we multiplied the observed RIXS 
intensity by $(1+\tan(\theta))$ to correct for absorption so that RIXS intensities at different ${\rm\bf q}$-positions can be directly compared.
All measurements were done at $T \sim 10$~K.

\begin{figure}
\includegraphics[width=8cm]{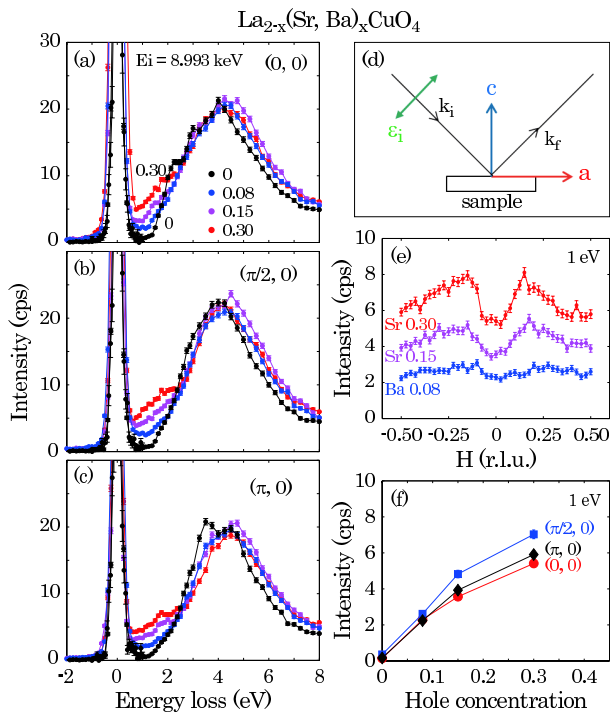}
\caption{(Color online) Figures (a), (b) and (c) show RIXS spectra at 
$(0, 0)$, $(\pi/2, 0)$ and $(\pi, 0)$, respectively, for the Ba0.08, Sr0.15, 
and Sr0.30 samples, measured with $E_i = 8.993$~keV.  The RIXS intensity 
below 3~eV increases monotonically with doping.  (d) Scattering geometry of 
the present study.  (e) The RIXS intensity profiles at energy transfer of 
1~eV along the $H$ direction for the three samples. (f) Hole concentration 
dependence of the RIXS intensity at 1~eV at $(0, 0)$, $(\pi/2, 0)$ and 
$(\pi, 0)$.}
\end{figure}

\section{Results}

\subsection{Measurements with $E_{i}=8.993$~keV}

Our previous measurements of stripe-ordered LBCO $x=0.125$ and 
LSCO $x=0.12$ show an enhancement of RIXS intensity at $\sim 1$~eV 
near $(\pm\pi/2, 0)$ and $(0, \pm\pi/2)$, which corresponds to the 
stripe-${\rm\bf q}$ positions.~\cite{Wakimoto2009}  
In order to test if this is a unique feature 
of the striped phase, we have carried out systematic measurements 
with various dopings with incident photon energy $E_{i}=8.993$~keV.

Figure~2~(a)-(c) shows RIXS spectra for undoped LCO, LBCO x=0.08, 
LSCO x=0.15, 
and LSCO x=0.30, measured at the $(0,0)$, $(\pi/2,0)$, and $(\pi,0)$ 
positions.  It is clearly shown that the RIXS intensity below 
$\sim 2$~eV increases with doping, as summarized 
in Fig.~2~(f) where RIXS intensity at 1~eV is plotted as a function of 
doping.  
Cu-$K$ edge RIXS results by Ellis {\it et al.}~\cite{Ellis2011} 
using a $\epsilon_i // c$ ($\sigma$-polarization) set-up 
show that the increase of the RIXS intensity 
below $~\sim 2$~eV coincides with the decrease of the shoulder of the 
CT peak near 3~eV.  This behavior is not pronounced in the present results probably because of different polarization 
conditions, since in our scattering geometry we observe $\pi$-polarization signals coincidentally.

To further complicate matters, the polarization dependence 
of the CT spectra in the carrier doped samples has not been fully understood yet.  
In this study we focus on the RIXS intensity below the CT gap, which 
appears to be qualitatively consistent between different polarization 
conditions.

\begin{figure}
\includegraphics[width=8cm]{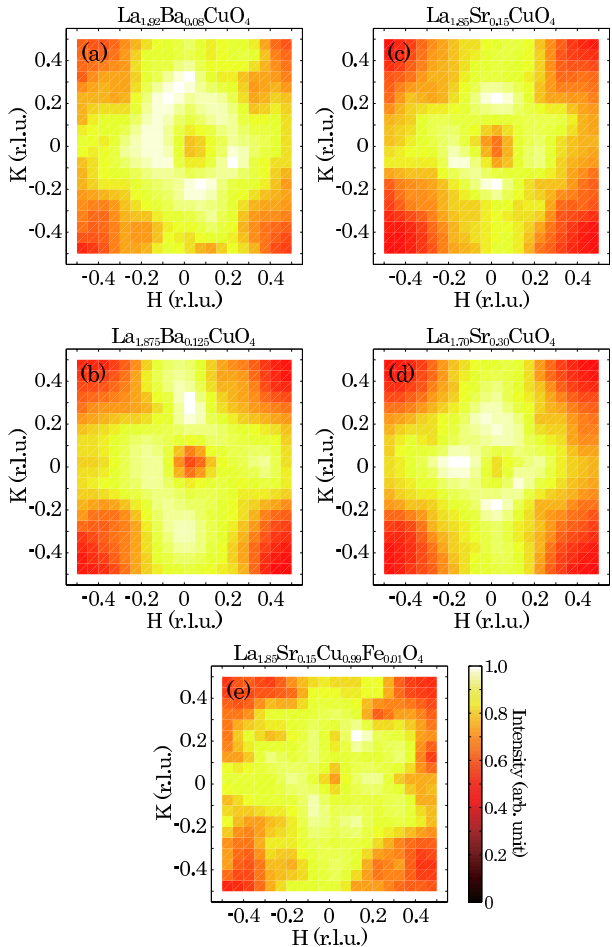}
\caption{(Color online) Contour plots of RIXS intensity at 1~eV for 
(a) La$_{1.92}$Ba$_{0.08}$CuO$_{4}$, (b) La$_{1.875}$Ba$_{0.125}$CuO$_{4}$, 
(c) La$_{1.85}$Sr$_{0.15}$CuO$_{4}$, (d) La$_{1.70}$Sr$_{0.30}$CuO$_{4}$, 
and (e) La$_{1.85}$Sr$_{0.15}$Cu$_{0.99}$Fe$_{0.01}$O$_{4}$.  
The intensity in each figure is corrected by an absorption factor 
$1+\tan(\theta)$, and normalized to the maximum intensity in each plot.}
\end{figure}

Figure~2~(e) shows RIXS profiles along $(H,0)$-direction at 1~eV for 
LBCO $x=0.08$, LSCO $x=0.15$, and LSCO $x=0.30$.  
As doping increases, the RIXS intensity 
increases and apparently a peak grows near $H=\pm0.2$, which corresponds to 
$(\pm0.4 \pi,0)$.  Previously we 
report that the stripe ordered samples have an intensity maximum at 
$(\pi/2, 0)$.~\cite{Wakimoto2009}  
However, those measurements were performed only for limited 
concentrations and at limited ${\rm\bf q}$ positions.  
Now, from the present detailed scan, it turns out that the maxima appear 
at the $(\pm0.4 \pi,0)$ positions, and this structure appears 
not only in the stripe-ordered $1/8$ but also in the overdoped 
sample.

With more careful examination of Fig.~2~(e), small humps near $H=\pm0.2$ also 
appear even in the LBCO $x=0.08$ sample.
In order to test if the maximum structure is a common feature within the hole-doped 
214 system, we performed $(H,K)$ mesh scans at $1$~eV for various 
concentrations and summarized the RIXS intensity maps in Fig.~3.  
Here the intensity is normalized to the maximum intensity in each mesh scan. 
It is seen that a ring-shaped intensity distribution resulting in  
intensity maxima near $(\pm0.4 \pi,0)$ and $(0,\pm0.4 \pi)$ appears in all LBCO and 
LSCO samples (Fig.~3~(a) to (d)).  
The Fe-doped LSCO data (Fig.~3~(e)) was obtained at 9-ID B by integrating 
the intensity in the energy range of $1\pm0.2$~eV.  
The ring-shaped structure also appears in this data, and therefore we can 
exclude spectrometer related extrinsic origin of the ring structure.  
Among these materials, LBCO $x=0.125$ and Fe-doped LSCO have 
charge-density-wave (CDW) orders: 
the former shows the magnetic and charge stripe orders below 
$\sim 50$~K,~\cite{Fujita2004} and the latter shows spin-density-wave and 
CDW orders below 20~K.
Thus, the ring structure is indeed a universal feature in the RIXS spectra 
measured with $E_i=8.993$~keV within the hole-doped 214 system regardless 
of the charge-stripe/CDW order.

\begin{figure}
\includegraphics[width=8cm]{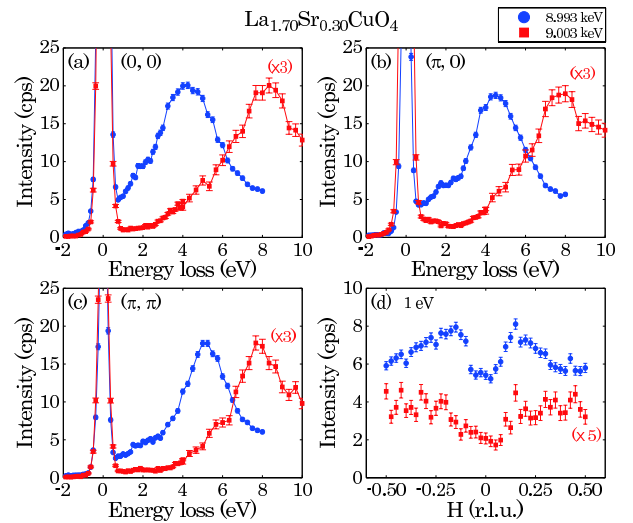}
\caption{(Color online) Comparison between RIXS spectra of 
La$_{1.70}$Sr$_{0.30}$CuO$_{4}$ measured with $E_i = 8.993$~keV 
(circles) and $9.003$~keV (squares) at (a) $(0, 0)$, (b) $(\pi, 0)$ 
and (c) $(\pi, \pi)$.  (d) compares $H$-scan profiles of RIXS intensity 
at 1~eV.  In all figures, the data with $E_i = 9.003$~keV are 
multiplied by 3 or 5, for clarity.}
\end{figure}

\subsection{Measurements with $E_{i}=9.003$~keV}

The RIXS study of the electron-doped system Nd$_{1.85}$Ce$_{0.15}$CuO$_{4}$ 
has revealed a dispersive intra-band charge excitation inside the CT-gap 
which is consistent with the dynamical charge-charge correlation 
function $N(q,\omega)$.~\cite{Ishii2005a}
To date, however, such intra-band excitation has not been identified in the
hole doped systems.  We have performed RIXS measurements for LSCO $x=0.30$ 
and overdoped Bi2201 with $E_i=9.003$~keV to detect such excitation.
We chose these compounds for since RIXS intensity 
below the CT gap increases with doping, and therefore the 
system becomes more metallic.  
The present overdoped Bi2201 sample shows no superconductivity down to 
2~K.  We estimate that the hole concentration of this sample is more 
than 0.2/Cu, after the phase diagram of Pb-doped Bi2201 by Kudo 
{\it et al.}~\cite{Kudo2009}

\begin{figure}
\includegraphics[width=6cm]{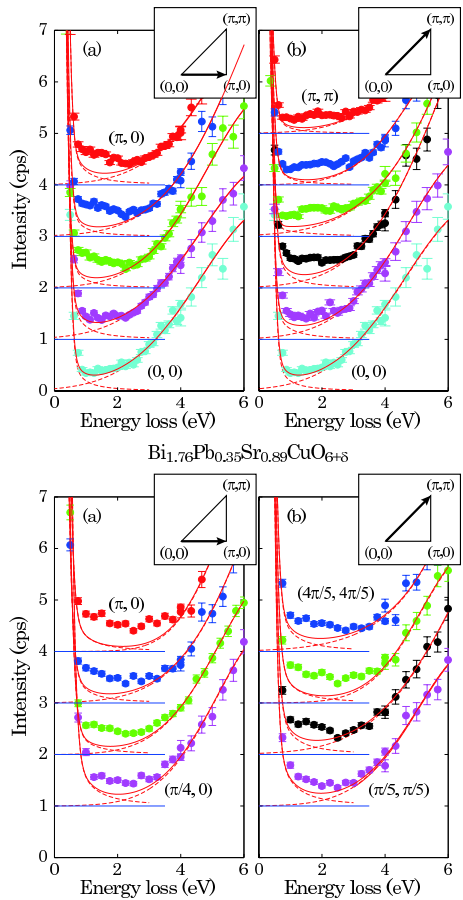}
\caption{(Color online) $q$-dependence of the RIXS spectra of 
La$_{1.70}$Sr$_{0.30}$CuO$_{4}$ measured with $E_i = 8.993$~keV 
along (a) the $(0, 0)$ to $(\pi, 0)$ direction and (b) the $(0, 0)$ 
to $(\pi, \pi)$ direction.  Each spectrum is shifted by 1 cps for clarity.  
Dashed and solid lines indicate fits to the elastic component, the molecular orbital 
excitation centered at 8~eV, and their sum respectively.  Figures (c) and (d) 
are analogous plots for Bi$_{1.76}$Pb$_{0.35}$Sr$_{1.89}$CuO$_{6+\delta}$.
Data at $(0, 0)$ and $(\pi, \pi)$ are missing due to contamination by 
large elastic components}
\end{figure}

Figures 4~(a)-(c) show RIXS spectra of LSCO $x=0.30$ measured at $E_i=8.993$ 
and $9.003$~keV.  The CT peak in the $8.993$~keV data, which disperses 
from 4 to 5~eV, is replaced by a peak at 8~eV in the 9.003~keV data.  
This non-dispersive excitation can be attributed to a molecular-orbital 
(MO) excitation.~\cite{Kim2004}
Although the overall RIXS intensity is smaller, the 9.003~keV data is 
preferable for evaluation of the intra-band excitation in the energy range below 
4~eV since the MO peak is located at higher energy than the CT peak in the 
8.993~keV data and the peak tail has a smaller effect in this energy range.
Figure 4~(d) shows the $H$-scan profiles measured at 1~eV energy transfer.  A remarkable difference between the two different $E_i$ profiles is 
that the maximum structure at $H=\pm0.2$ is not clear in the 
9.003~keV data.  However, both show that the intensity has a 
minimum at the $\Gamma$ point.

\begin{figure}
\includegraphics[width=8cm]{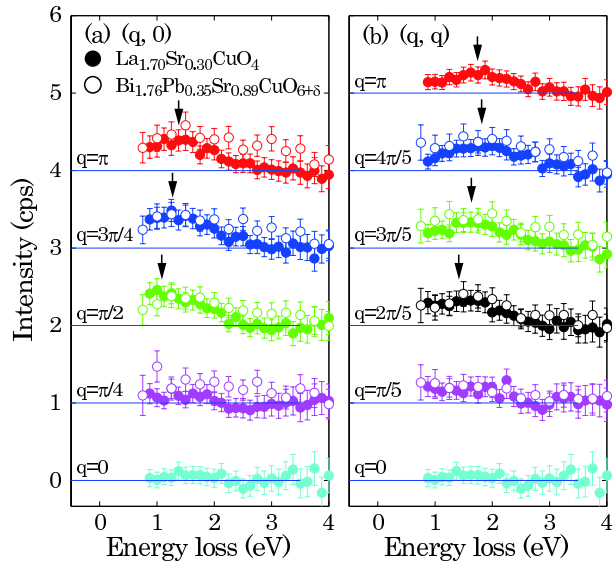}
\caption{(Color online) RIXS intensity after the subtraction of the 
elastic and molecular orbital components for 
La$_{1.70}$Sr$_{0.30}$CuO$_{4}$ (closed circles) and 
Bi$_{1.76}$Pb$_{0.35}$Sr$_{1.89}$CuO$_{6+\delta}$ (open circles).  
(a) $q$-dependence along the $(0, 0)$ to $(\pi, 0)$ direction.  
(b) $q$-dependence along the $(0, 0)$ to $(\pi, \pi)$ direction.
Arrows indicate center of mass of the remnant intensity of LSCO $x=0.30$
determined by fitting to a Gaussian function.}
\end{figure}

To extract intra-band excitation in the energy range below 4~eV, we have 
subtracted the elastic and MO peak components from the 9.003~keV spectra.  
The elastic tail was estimated from the anti-Stokes region.  The MO peak 
tail was estimated by fitting the MO peak to a Gaussian function for the 
energy range above 4~eV.  
This is the same method used to extract the intra-band excitation described 
in Ref.~\onlinecite{Ishii2005a}.
Figure 5 shows the variation of spectra for LSCO $x=0.30$ ((a) and (b)) and 
Bi2201 ((c) and (d)) as ${\rm\bf q}$ changes from $(0,0)$ to $(\pi,0)$ and from 
$(0,0)$ to $(\pi,\pi)$.
The estimated elastic tail and MO peak components are indicated by dashed 
lines and their sum is shown as solid lines.  
For the data of LSCO $x=0.30$ in Figs.~5~(a) and (b), additional 
intensity above the solid line below 3~eV becomes observable as the 
${\rm\bf q}$-position departs from $(0,0)$, while it is difficult 
to distinguish any additional intensity above the elastic tail and the MO 
components at $(0,0)$.  
A similar behavior can be found in the data of Bi2201 in Figs.~5~(a) and (b).  
Here, the data at $(0,0)$ and $(\pi,\pi)$ are not shown since 
they are highly contaminated by a large elastic component, making the evaluation 
of the intra-band excitation for these points largely ambiguous.
Nevertheless, for the remaining profiles, additional intensity above 
the solid lines is clearly recognizable.

Remnant intensities after the subtraction of the elastic and 
MO components are summarized in Figs.~6 and 7.
Figure~6 plots the remnant intensities at the ${\rm\bf q}$-positions corresponding 
to those in Fig.~5.  Data below 0.8~eV are removed since they are largely 
ambiguous due to the rapid increase of the elastic component in this energy region.  
The remnant intensities are nearly identical 
for the two samples, LSCO $x=0.30$ and overdoped Bi2201, suggesting that 
the observed feature is common in hole-doped systems.  
Arrows indicate the center of mass of the remnant intensity of LSCO $x=0.30$ 
determined by fitting the data to a Gaussian function.  The center of mass 
tends to move to higher energy as ${\rm\bf q}$ increases, thus showing dispersion.  
Figure 7 shows a color map of the remnant intensities of LSCO $x=0.30$.  
This intensity distribution remarkably resembles that of the dispersive 
intra-band excitation in NCCO.~\cite{Ishii2005a}
Here we indicate the center of mass positions by circles, together with those 
for NCCO indicated by squares, adopted from Ref.~\onlinecite{Ishii2005a}.  
The intra-band excitation we have observed has a similar dispersion to that of 
NCCO, and thus, it is likely to have the same origin.

\begin{figure}
\includegraphics[width=8cm]{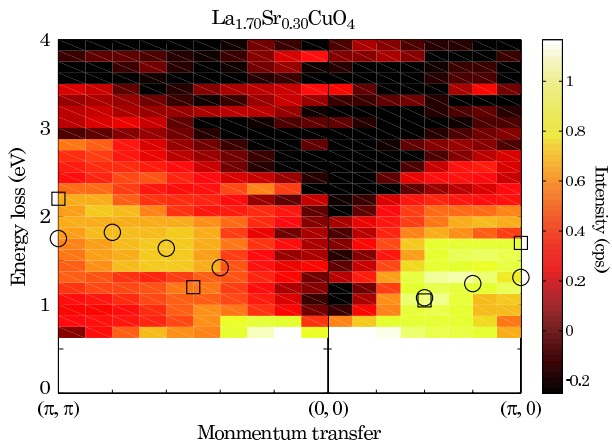}
\caption{(Color online) A $q-E$ contour plot of the RIXS intensity 
after the subtraction of the elastic and molecular orbital 
components for La$_{1.70}$Sr$_{0.30}$CuO$_{4}$.
Circles indicate the positions of the center of mass for LSCO $x=0.30$ 
shown by arrows in Fig. 6, while squares indicate those for NCCO 
adopted from Ref.~\onlinecite{Ishii2005a}. 
}
\end{figure}

\section{Discussion}


Judging by our measurements with $E_i=9.003$~keV, we have identified a dispersive 
intra-band excitation in the hole doped systems, similar to that of the 
electron doped systems.
The intra-band excitation in NCCO was attributed to the dynamical charge-charge 
correlation function $N(q,\omega)$. Similarly, it is 
natural to consider that the dispersive intra-band excitation we 
observe is the dynamical charge correlation function $N(q,\omega)$, therefore concluding 
that the electron and hole doped systems have similar $N(q,\omega)$. 
In fact, theoretical calculations indicate that $N(q,\omega)$ has 
qualitatively similar ${\rm\bf q}$-dependence in both electron- and 
hole-doped systems.\cite{Tohyama_pri,Jia2011}

It has been theoretically demonstrated that the indirect RIXS 
cross-section is linear with the dynamical charge correlation 
function at the limit where the short-lived, local core-hole 
potential is either weak or strong.\cite{Brink2006,Ament2007}  
On the other hand, Ishii {\it et al}.\cite{Ishii2005a} suggested 
that the intra-band excitation is pronounced in the RIXS spectra 
when the core hole is created at the carrier-doped site, and the 
spectra are similar to $N(q,\omega)$.  
Recently, Jia {\it et al}.\cite{Jia2011} examined detailed 
$E_i$-dependences of the Cu-$K$ edge RIXS spectra for both hole 
and electron doped cuprates numerically and confirmed that the
$N(q,\omega)$-type intra-band excitation becomes dominant in the RIXS 
spectra when the incident photon energy is tuned to the resonance 
energy of Cu sites where doped electrons/holes are located.  
In the electron doped case, this incident energy is located slightly 
below the well-screened condition, while in the hole doped case 
it is located slightly above the poorly-screened condition.  This is 
consistent with the experimental fact in the present study that 
the $N(q,\omega)$ type intra-band excitation of the hole-doped 
system was observed with the resonance energy near the 
poorly-screened condition, namely 9.003~keV.  
At these incident energies the core hole is created at the 
carrier-doped site and its potential does not perturb the 
valence electrons significantly, therefore this resonant condition 
is close to the aforementioned weak limit of the core-hole potential. 
These agreements with theoretical studies support our assertion that the dispersive 
intra-band excitation we observe arises from $N(q,\omega)$.

On the other hand, we also note some disagreements. In our measurements, the RIXS 
intensity around 1~eV decreases monotonically with increasing $E_i$ from 
8.993 to 9.003~keV, while the main CT/MO peak shifts to higher energy.  
The incident energy of 9.003~keV we used provide a compromise 
between the disadvantage of decreasing intensity and the advantage 
of decreasing MO peak tail.
However, theoretical calculations predict a monotonic increase of the 
$N(q,\omega)$ contribution when $E_i$ changes from 8.993 to 
9.003~keV,~\cite{Jia2011} which disagrees with our observations.  
This implies that the RIXS intensity inside the CT gap in the 8.993~keV 
data contains appreciable contribution from another type of excitation.  
In this regard, Ellis {\it et al.} have hypothesized the excitation near 
2~eV measured with 8.993~keV to be an inter-band excitation from a 
``low-lying'' band below the Fermi energy $E_F$ to the Zhang-Rice band.  
The incident and final polarization conditions of x-ray photons may also affect 
the $E_i$ dependence of the RIXS intensity.  However, RIXS spectra 
with $E_i = 8.993$~keV taken with different polarization conditions, namely
the present data and those reported in Ref.~\onlinecite{Ellis2011}, 
are very similar.  
In addition, Chabot-Couture {\it et al}.~\cite{Chabot2010} reported 
weak polarization dependence of the Cu-$K$ edge RIXS spectra of 
Nd$_2$CuO$_4$.  Therefore, the polarization-dependence in cuprates is 
likely to be weak as well.  Though polarization dependence is not taken into 
account in the theoretical calculations mentioned, the disagreement regarding the 
enhanced intensity of the intra-band region at $E_i = 8.993$~keV 
between theory and our experiments can be hardly attributed 
to polarization effects.  More detailed investigations are 
needed to understand the $E_i$ dependence of the intra-band intensity.


A characteristic feature in the 8.993~keV data is that the RIXS 
intensity profiles at 1~eV have maxima at $(\pm0.4 \pi,0)$ in the 
entire hole-doping range.  
Previously we have hypothesized that this feature is related to 
the stripe ordered state,\cite{Wakimoto2009} 
but as we show here the maximum structure appears in all 
samples regardless of the existence of charge stripe order.  
Thus, it is unlikely that the maximum structure is originating directly 
from the stripe ordered phase.  
We have reported the enhancement of the Ni-K edge RIXS intensity 
near 0.5~eV at the stripe-$q$ positions in stripe ordered 
La$_{5/3}$Sr$_{1/3}$NiO$_{4}$.\cite{Wakimoto2009}  However, 
Simonelli {\it et al}.\cite{Simonelli2010} have reported the 
absence of such enhancements by equally sensitive RIXS
measurements. Therefore, the stripe-originated charge excitation 
has not been confirmed in the nickelate case, either.

So far, however, we do not have enough experimental and theoretical 
clues to completely rule out the possibility that 
the maximum behavior in the cuprates 
arises from charge instability towards charge stripes.  
More RIXS studies of charge-ordered systems are
necessary to elucidate the nature of charge fluctuations and charge order.


\section{Summary}

We have carried out RIXS measurements near the Cu-$K$ absorption edge 
using hole-doped LSCO, LBCO and Bi2201, covering a wide 
hole-concentration range from undoped to overdoped, focusing 
on the charge excitations below the CT gap.  
Measurements with $E_i=8.993$~keV, which involves the 
well-screened intermediate state, have shown that the RIXS 
intensity near 1~eV has maxima at $(\pm0.4 \pi,0)$ for all doped samples 
regardless of the presence of charge stripe order.  
This means that the maximum structure is not directly related to 
the charge stripe ordered state.
Measurements with $E_i=9.003$~keV for overdoped LSCO and Bi2201 
have revealed the existence of a dispersive intra-band excitation 
which is similar to that in the electron-doped NCCO.  
This excitation can be attributed to the dynamical charge-charge 
correlation function, which is then very similar for both the electron 
and hole-doped systems.

\begin{acknowledgments}

We thank K. Tsutsui, T. Tohyama, D. Ellis, and T. Devereaux,
for invaluable discussion.
This work was supported by Grant-in-Aids from The Ministry of Education, 
Culture, Sports, Science and Technology (MEXT), Japan (No.22604007).
The synchrotron radiation experiments were performed at the BL11XU at 
SPring-8 with the approval of the Japan Synchrotron Radiation Research 
Institute (JASRI) (Proposal No. 2009A3502, 2009B3502, 2010A3502, 2011B3502).
Use of the Advanced Photon Source, an Office of Science User Facility 
operated for the U.S. Department of Energy (DOE) Office of Science by
Argonne National Laboratory, was supported by the U.S. DOE under 
Contract No. DE-AC02-06CH11357.

\end{acknowledgments}

\end{document}